\documentclass[aps,prr,reprint, amsmath, amssymb,superscriptaddress]{revtex4-1}

\pdfoutput=1
\usepackage[utf8]{inputenc}
\usepackage{bm}
\usepackage[english]{babel}
\usepackage[T1]{fontenc}
\usepackage{mathrsfs}
\usepackage[retainorgcmds]{IEEEtrantools}
\usepackage{amsmath}
\usepackage{amssymb}
\usepackage{color}
\usepackage{amsfonts}
\usepackage{times,txfonts}
\usepackage{nicefrac}
\usepackage[colorlinks=true,linkcolor=blue,urlcolor=blue,citecolor=blue,pdfusetitle]{hyperref}
\usepackage{mathtools}

\usepackage{float}
\usepackage{epsfig} 
\usepackage{subfigure}
\graphicspath{{images/}}

\usepackage{physics}

\usepackage{lipsum}
\usepackage{enumerate}

\usepackage[dvipsnames]{xcolor}
\usepackage{accents}

\begin{document}

\title{Quantum thermodynamics with fast driving and strong coupling via the mesoscopic leads approach}

\author{Artur M. Lacerda}
\affiliation{Department of Physics, Trinity College Dublin, Dublin 2, Ireland}
\author{Archak Purkayastha}
\email{archak.p@phys.au.dk}
\affiliation{Department of Physics, Trinity College Dublin, Dublin 2, Ireland}
\affiliation{Center for Complex Quantum Systems, Department of Physics and Astronomy, Aarhus University, Ny Munkegade 120, DK-8000 Aarhus C, Denmark.}
\affiliation{Department of Physics, Indian Institute of Technology, Hyderabad 502284, India}
\author{Michael Kewming}
\affiliation{Department of Physics, Trinity College Dublin, Dublin 2, Ireland}
\author{Gabriel T. Landi}
\affiliation{Department of Physics, Trinity College Dublin, Dublin 2, Ireland}
\affiliation{Instituto de Física da Universidade de São Paulo, 05314-970 São Paulo, Brazil}
\author{John Goold}
\affiliation{Department of Physics, Trinity College Dublin, Dublin 2, Ireland}
\begin{abstract}

Understanding the thermodynamics of driven quantum systems strongly coupled to thermal baths is a central focus of quantum thermodynamics and mesoscopic physics. A variety of different methodological approaches exist in the literature, all with their own advantages and disadvantages. The mesoscopic leads approach was recently generalised to steady state thermal machines and has the ability to replicate Landauer-Büttiker theory in the non-interacting limit. In this approach a set of discretised lead modes, each locally damped, provide a Markovian embedding for the baths.  
In this work we further generalise this approach to incorporate an arbitrary time dependence in the system Hamiltonian. Following a careful discussion of the calculation of thermodynamic quantities we illustrate the power of our approach by studying several driven mesoscopic examples coupled to finite temperature fermionic baths, replicating known results in various limits. In the case of a driven non interacting quantum dot we show how fast driving can be used to induce heat rectification. 
\end{abstract}

\maketitle

\section{Introduction}
Thermodynamics is a pillar of physical, chemical and biological sciences primarily due to the unique synergy it offers between the fundamental and pragmatic. It tells us that heat and work are just two different forms of energy (the first law), and it sets constraints on the heat generated by mechanical operations that we are allowed to perform  (the second law)~\cite{callen}. 
Importantly, the notion of thermal equilibrium is central for defining the state functions of conventional thermodynamics. 
However, when we move towards the microscopic domain, understanding thermodynamics far from equilibrium becomes increasingly important~\cite{de2013non,jarzynski2011equalities,
seifert2012stochastic,seifert2019stochastic}.

In particular, as we go towards the advent of quantum devices across various platforms like atoms and ions in optical traps \cite{Monroe_2013}, molecular junctions \cite{Aradhya_2013}, quantum dots \cite{Kagan_2021}, hybrid semiconductor-superconductor circuits \cite{Bukard_2020,Clerk_2020}, nitrogen-vacancy centers in diamonds \cite{Doherty_2013} etc., it has become necessary to understand the energy and the entropy cost of such applications \cite{Arrachea_2022,Auffeves_2021}. These devices are inherently noisy because complete isolation from surrounding environment is neither possible at such small length scales, nor desired in many applications. In fact, in some applications, coupling to multiple baths is desired, for example, to induce a voltage bias (for example, \cite{Liu_2015,Stehlik_2016,Wen_2020}).  Moreover, the working principle of these devices often require control via external driving, modelled by a time dependent Hamiltonian \cite{Koch_2019,James_2021,adiabat_engine}. To describe and understand the energy and the entropy cost of such devices, one needs (a) a consistent extension of the standard notion thermodynamics to externally driven microscopic quantum systems coupled to multiple macroscopic baths, (b) a prescription to calculate the required thermodynamic quantities. In this paper, we show that the so-called mesoscopic leads approach (also known by other names, such as pseudomodes approach and driven Liouville von-Neumann equation) \cite{Imamoglu_1994,Garraway_1997a,Garraway_1997b,Sanchez_2006,Subotnik_2009,Dzhioev_2011,
Ajisaka_2012,Ajisaka_2013,Arrigoni_2013,Dorda_2014,Chen_2014,Zelovich_2014,
Dorda_2015,Hod_2016,Gruss_2016,Schwarz_2016,Dorda_2017,Elenewski_2017,
Gruss_2017,Zelovich_2017,Tamascelli_2018,Lemmer_2018,Oz_2019,Chen_Galperin_2019,
Zwolak_2020,Wojtowicz_2020,Chiang_2020,Brenes2020,Lotem_2020,Fugger_2020,
Wojtowicz_2021,Elenewski_2021} achieves these goals, providing a numerically efficient formalism applicable far beyond regimes accessible by other approaches.

Most standard formalisms to describe such open quantum systems rely on weak system-bath couplings such that a Born-Markov approximation can be made, and an effective quantum master equation can be derived \cite{Breuer_book,Rivas_book}. Although usually analytically and numerically the simplest, various drawbacks of such approaches have been pointed out \cite{Walls1970,Novotny_2002,Wichterich_2007,Novotny_2010,
Rivas_2010,barranco_2014,Levy2014, archak_2016, Eastham_2016, Hofer_2017,Gonzalez_2017,Mitchison_2018, Cattaneo_2019,Hartmann_2020_1,konopik_2020local, Scali_2021,Archak_2021}, and it has been shown that no such approximation can generically accurately describe the long-time state of the system in presence of multiple baths \cite{Archak_2021}. Moreover, in many applications, like those involving quantum dots and molecular junctions, experimentally the system-bath couplings are often not desired to be weak \cite{Aradhya_2013,Bukard_2020,Clerk_2020,Kagan_2021}. The presence of external driving makes the situation even more complicated, since driving can introduce further non-Markovian effects (see for example, \cite{Schnell_2020,Schnell_2021}). 

In the non-Markovian regime, the most widely used formalism are those of non-equilibrium Green's functions (NEGF) \cite{Jauho_book} and Schwinger-Keldysh path integrals \cite{Kamenev_book}. Other formalisms like microscopically derived quantum Langevin equations \cite{Cortes_1985,Ford_1988,Kac_1981,Dhar_2003,Dhar_Roy_2006}, and the scattering approach \cite{Moskalets_book,Datta_1987,Arrachea_2006,Lesovik_2011,Brandner_2020,Potanina_2021} are closely related to these approaches. In absence of many-body interactions (i.e, higher than quadratic terms in fermionic or bosonic creation and annihilation operators in Hamiltonian), these approaches can efficiently give exact results for long-time steady state if there is no external driving. Treating systems with many-body interaction using these approaches requires existence of a small parameter, such that a diagrammatic expansion is possible \cite{Jauho_book, Kamenev_book}. Further, in presence of an external driving, often even quadratic Hamiltonians cannot be treated exactly. Assuming that the drive is periodic, one has to often rely on perturbative expansions either in frequency or in strength of the drive (for example, \cite{Ludovico_2016,Bruch_2018,RieraCampeny2019,Brandner_2020}).  In experimental situations, however, one may not be just limited to such regimes  (for example, \cite{Stehlik_2016}). Moreover, even in cases where these approaches efficiently give the long-time behavior, obtaining the full behavior from short-times to long-times is usually quite difficult.

The mesoscopic leads approach is an alternate way to obtain the dynamics of the system in strong-coupling non-Markovian regime \cite{Imamoglu_1994,Garraway_1997a,Garraway_1997b,Sanchez_2006,Subotnik_2009,Dzhioev_2011, Ajisaka_2012,Ajisaka_2013,Arrigoni_2013,Dorda_2014,Chen_2014,Zelovich_2014,
Dorda_2015,Hod_2016,Gruss_2016,Schwarz_2016,Dorda_2017,Elenewski_2017,
Gruss_2017,Zelovich_2017,Tamascelli_2018,Lemmer_2018,Oz_2019,Chen_Galperin_2019,
Zwolak_2020,Wojtowicz_2020,Chiang_2020,Brenes2020,Lotem_2020,Fugger_2020,
Wojtowicz_2021,Elenewski_2021}. In this approach, the macroscopic baths are systematically approximated by a finite number of damped modes, which we call leads. Initially introduced for bosonic baths \cite{Imamoglu_1994,Garraway_1997a,Garraway_1997b}, this approach has been subsequently extensively used to study quantum transport in fermionic set-ups also \cite{,Sanchez_2006,Subotnik_2009,Dzhioev_2011,
Ajisaka_2012,Ajisaka_2013,Arrigoni_2013,Dorda_2014,Chen_2014,Zelovich_2014,
Dorda_2015,Hod_2016,Gruss_2016,Schwarz_2016,Dorda_2017,Elenewski_2017,
Gruss_2017,Zelovich_2017,Chen_Galperin_2019,Oz_2019,
Zwolak_2020,Wojtowicz_2020,Chiang_2020,Brenes2020,Lotem_2020,Fugger_2020,
Wojtowicz_2021,Elenewski_2021}. When combined with tensor network techniques,  it has been recently shown that it is possible to completely non-perturbatively obtain energy and particle currents in non-equilibrium steady states (NESS) of interacting quantum many-body systems, and thereby the thermodynamics at NESS, using this approach \cite{Brenes2020}. Related approaches have been developed to describe impurity models at and beyond Kondo regimes \cite{Arrigoni_2013,Dorda_2014,Dorda_2015,Dorda_2017,Lotem_2020,Fugger_2020}. However, most existing investigations were limited to time-independent Hamiltonians, with only few considering time-dependent system Hamiltonians \cite{Chen_2014,Oz_2019}. 

Formalisms to describe thermodynamics of externally driven systems coupled to baths have been investigated using various approaches and approximations \cite{Esposito_2015_1,Brandner_2016,Bruch_2018,Brandner_2020,
Potanina_2021,Dann_2021}, each with its own merits and drawbacks, and applied to simple set-ups \cite{Esposito_2015_2,Ludovico_2016,Ludovico_2016_2,Bruch_2016,Katz_2016,Haughian_2018,
RieraCampeny2019,Oz_2019}. But, to our knowledge, thermodynamics with time dependent drive in the mesoscopic leads approach has only been previously considered in the case of a driven resonant level model, where the thermodynamics has been described only in the regime of slow driving \cite{Oz_2019}. Here, we extend the mesoscopic leads approach to arbitrary system Hamiltonians in the presence of arbitrary time dependent driving. We start from a microscopic description which is common to all standard approaches to open quantum systems (Sec.~\ref{subsec:open_quantum_systems}). Following existing approaches to thermodynamics of quantum systems \cite{Esposito2009a, Reeb_2014, Landi_2021, Strasberg_tutorial_2021}, we then give a careful and detailed discussion of the quantities that are required to be calculated in order to describe the thermodynamics in the presence of an arbitrary external drive and coupling to multiple baths (Sec.~\ref{subsec:thermodynamics}). Next, we discuss how the mesoscopic leads approach can be systematically derived from the microscopic description (Sec.~\ref{subsec:mesoscopic_leads_derivation}), and show that all quantities required for the description of thermodynamics can be obtained via this approach (Sec.~\ref{subsec:thermodynamics_mesoscopic_leads}). Then, we lay down an elegant formalism for quadratic Hamiltonians, which holds for arbitrary driving in the system and arbitrary strengths of system-bath couplings, and can easily describe dynamics and thermodynamics at all times. This therefore provides a simple way to access cases beyond both NEGF and standard weak-coupling quantum master equation descriptions (Secs.~\ref{subsec:lyapunov_eq}, \ref{subsec:floquet}).  We apply the formalism to two examples: a driven resonant level (Sec.~\ref{sec:resonant_level}), and a two-site fermionic system with one site being externally periodically driven (Sec.~\ref{sec:rectification}). We use the former for benchmarking our approach. In the later, motivated by \cite{RieraCampeny2019} we investigate rectification of energy currents. In particular, we obtain the entropy cost of energy rectification as a function of drive frequency. This would be difficult to obtain using most other approaches. Finally, we conclude by summarizing and giving the outlook (Sec.~\ref{sec:conclusion}).

\section{\label{sec:thermodynamics_oqs}Thermodynamics of driven open quantum systems}

\subsection{Microscopic description of open quantum
systems} 
\label{subsec:open_quantum_systems}

Let us begin by describing a general system configuration; a microscopic system of interest coupled to multiple macroscopic baths.
The former is controlled by some external protocol, mathematically described by a time-dependent Hamiltonian $\hat{H}_\text{S}(t)$. 
The Hamiltonian describing the coupling between the baths and the central system is thus
\begin{equation}
\label{eq:Hamiltonian}
    \hat{H}_{\text{tot}} = \hat{H}_{\text{S}}(t) + \sum_{\alpha}\qty(\hat{H}_\alpha + \hat{H}_{\text{S}\alpha}),
\end{equation}
where $\alpha$ labels the different baths, $\hat{H}_{\alpha}$ is the free Hamiltonian of the baths, and $\hat{H}_{S\alpha}$ describes the coupling between the baths and the system. We assume that initial state of the system is in an uncorrelated product state with the baths.
Furthermore, we assume that each bath begins in a thermal state at inverse temperature $\beta_\alpha$ and chemical potential $\mu_\alpha$.  Thus the joint state is
\begin{equation}
\label{eq:initial_state}
    \hat{\rho}_{\text{tot}}(0) = \hat{\rho}_{\text{S}}(0)  \prod_\alpha \hat{\rho}_{\alpha}^\text{th}, \qq{where} \hat{\rho}_{\alpha}^\text{th} = \frac{e^{-\beta_\alpha(\hat{H}_\alpha - \mu_\alpha \hat{N}_\alpha)}}{Z_\alpha}.
\end{equation}
where $\hat{N}_{\alpha}$ is the number operator of the $\alpha$th bath and $Z_{\alpha} = \Tr[e^{-\beta_\alpha(\hat{H}_{\alpha} - \mu_\alpha\hat{N}_{\alpha})}]$ is the corresponding partition function. 
Additionally, we assume that the bath Hamiltonian conserves the number of particles, i.e., $\comm{\hat{H}_\alpha}{\hat{N}_\alpha} = 0$. Starting from the initial state Eq.~\eqref{eq:initial_state}, the entire system is left to evolve unitarily under the Hamiltonian $\hat{H}_{\text{tot}}$, according to $\dd \hat{\rho}_{\text{tot}}(t)/ \dd t = -i[\hat{H}_{\text{tot}}(t),\hat{\rho}_{\text{tot}}]$ (we set $\hbar = 1$), so at time $t$ the total density-matrix is
\begin{equation}
    \hat{\rho}_{\text{tot}}(t) = \hat{U}(t)\,\hat{\rho}_{\text{tot}}(0)\hat{U}^\dagger(t),
\end{equation}
where $\hat{U}(t) = \mathcal{T}e^{-i\int_0^t \dd t' \hat{H}_{\text{tot}}(t')}$ is the time-evolution operator and $\mathcal{T}$ is the time-ordering operator. 
The evolution of the system of interest, $\hat{\rho}_{\text{S}}$, can be obtained by tracing out the the bath. 
We will denote $L_{\text{S}}$ and $L_{\text{B}}$ as the number of degrees of freedom (e.g., the number of sites in lattice descriptions) in the system and in each bath, respectively. 
Since the system is microscopic, $L_{\text{S}}$ is finite. Each bath, on the other hand, is by assumption macroscopic, thus $L_{\text{B}} \to \infty$. 
However, for mathematical rigorousness and to avoid unwanted divergences, 
this limit should be taken only after the bath degrees of freedom have been traced out. That is, 
\begin{equation}
    \hat{\rho}_{\text{S}}(t) = \lim_{L_{\text{B}} \to \infty} \Tr_{\text{B}}[\hat{\rho}_{\text{tot}}(t)],
\end{equation}
where $\Tr_{\text{B}}[\ldots]$ refers to trace over all bath degrees of freedom. The order of the operations is crucial, with the limit is taken after the trace. In order to avoid carrying cumbersome notation, we define the expected value of some generic operator $\hat{\mathcal{O}}$, which may be time-dependent, as
\begin{equation}
\label{eq:expectation_notation}
    \ev{\hat{\mathcal{O}}(t)} = \lim_{L_{\text{B}}\to\infty}\Tr[\hat{\mathcal{O}}(t)\hat{\rho}_{\text{tot}}(t)].
\end{equation}
In many cases, we will be interested in the long time behavior of the system which corresponds to
\begin{equation}
    \lim_{t \rightarrow \text{large}} \ev{\hat{\mathcal{O}}(t)} = \lim_{t \rightarrow \text{large}}\left( \lim_{L_{\text{B}}\to\infty}\Tr[\hat{\mathcal{O}}\hat{\rho}_{\text{tot}}(t)]\right).
\end{equation}
Once again, the order of limits cannot be interchanged. 

The above unitary picture of describing thermal baths is the starting point for microscopic derivations of all standard approaches to open quantum systems, such as non-equilibrium Green's functions (NEGF) \cite{Jauho_book}, Feynman-Vernon influence functional methods \cite{Kamenev_book}, quantum master equations \cite{Breuer_book} and quantum Langevin equations \cite{Cortes_1985,Ford_1988,Kac_1981,Dhar_2003,Dhar_Roy_2006}. The crucial point is the microscopic nature of the system and the macroscopic nature of the baths.
Given its microscopic description, the system is assumed to be accessible to microscopic measurements.
This is not true for the macroscopic baths whose microscopic details are not accessible. Instead, only knowledge of some of their macroscopic properties such as the temperatures, chemical potentials, and spectral functions are assumed to be known. 
Consequently, effective expressions for system observables and currents from baths are obtained after integrating out the baths. The various standard techniques of open quantum systems allow one to compute these, albeit at differing levels of approximation \cite{Jauho_book,Kamenev_book,Breuer_book}. 
With this microscopic setting in mind, we will now turn to describing the necessary thermodynamic quantities and how they can be studied in this context.

\subsection{Description of the thermodynamics}
\label{subsec:thermodynamics}
In this section, we will discuss the thermodynamic quantities in the setup described in the previous section. Our choice of definitions is completely self consistent, despite not being unique. A careful discussion of this point is given further in this section.
We focus on the average thermodynamic currents, such as work, heat and entropy production. 
The work performed by the external protocol is defined as the change in energy of the global set-up,
\begin{equation}
    \label{eq:ext_work}
    W_{\text{ext}}(t) = \ev{\hat{H}_{\text{tot}}(t)} - \ev{\hat{H}_{\text{tot}}(0)}.
\end{equation}
Under this sign convention, work is positive if done \textit{on} the system and negative if it is done \textit{by} the system. Since only the system Hamiltonian is time-dependent, the work can be re-written as follows,
\begin{equation}
    W_{\text{ext}}(t) = \int_0^t \dd t' \dv{\!\ev{\hat{H}_{\text{tot}}(t')}}{t'}
    =\int_0^t \dd t' \ev{\pdv{\hat{H}_\text{S}(t')}{t'}},
\end{equation}
where we used the fact that $\partial \hat{H}_{\text{tot}}/ \partial t  = \partial \hat{H}_{\text{S}}/ \partial t$.

We define the entropy production as Refs.~\cite{Esposito2009a, Reeb_2014, Landi_2021, Strasberg_tutorial_2021},
\begin{align}
\label{eq:ent_prod_relative_ent}
    \begin{split}
        \Sigma(t) &= \lim_{L_B \to \infty}D\left(\hat{\rho}_{\text{tot}}(t)~ \big\| ~ \hat{\rho}_{\text{S}}(t) \prod_\alpha \hat{\rho}_{\alpha}^\text{th}\right),
    \end{split}
\end{align}
where 
$D\left( \hat{\rho} ~ \middle\| ~ \hat{\sigma}\right) = \Tr[\hat{\rho}\log\hat{\rho} - \hat{\rho}\log\hat{\sigma}]$
is the quantum relative entropy between the density matrices $\hat{\rho}$ and $\hat{\sigma}$. 
Since $D\left( \hat{\rho} ~ \middle\| ~ \hat{\sigma}\right) \geq 0$, we have that $\Sigma(t) \geq 0$, so the second law of thermodynamics is naturally satisfied. 
Given this definition, one can recast the entropy production in terms of the dissipated heat and the change in the von-Neumann entropy of the system only  \cite{Esposito2009a, Landi_2021}
\begin{equation}
\label{eq:ent_prod_clausius}
    \Sigma(t)= \delta S_\text{S}(t) - \sum_\alpha \beta_\alpha Q_\alpha(t)\,,
\end{equation}
where $\delta S_{\text{S}}$ is the change in von-Neumann entropy of the system, $\delta S_{\text{S}} = S_{\text{S}}(t) - S_{\text{S}}(0)$,
with $S(t)=-{\rm Tr}(\hat{\rho} \log \hat{\rho})$ and $Q_\alpha$ is the heat dissipated into the $\alpha$-th bath given by
\begin{equation}
    \label{eq:heat_alpha}
     Q_\alpha(t) = \ev{\hat{H}_\alpha(t)} - \ev{\hat{H}_\alpha(0)} - \mu \qty(\ev{\hat{N}_\alpha(t)} - \ev{\hat{N}_\alpha(0)})\,.
\end{equation}
Apart from the work done by the external protocol, there is a chemical work due to the non-zero chemical potentials of the baths. Chemical work done by the system generates an increase in the number of particles within the baths,
\begin{equation}
\label{eq:chem_work}
    W_\text{chem}(t) = -\sum_\alpha \mu_\alpha \qty(\ev{\hat{N}_\alpha(t)} - \ev{\hat{N}_\alpha(0)})\,.
\end{equation}
The total work is therefore given by the sum of the external work Eq.~(\ref{eq:ext_work}) and the chemical work
\begin{align}
W(t)=W_\text{chem}(t)+W_{\text{ext}}(t).
\end{align}
The change in internal energy is defined by
\begin{equation}
    \delta U(t) = W(t) - \sum_\alpha Q_\alpha(t)\,,
\end{equation}
which corresponds to the first law of thermodynamics. 
By substituting in the definitions of both the total work, the dissipated heat, and the expression for $\hat{H}_{\rm tot}$ given by Eq.~\ref{eq:Hamiltonian}, one finds that the change in internal energy is 
\begin{equation}
    \label{eq:internal_energy}
    \delta U(t) = \ev{\hat{H}_{\text{S}}(t) + \sum_\alpha\hat{H}_{\text{S}\alpha}(t)} - \ev{\hat{H}_{\text{S}}(0) + \sum_\alpha\hat{H}_{\text{S}\alpha}(0)}.
\end{equation}
Intuitively, the change in the internal energy of the system only depends on those components of the Hamiltonian that relate to the system.

We also call attention to the limit $L_B \to \infty$ in Eq.~\eqref{eq:ent_prod_relative_ent}, which permits definitions of the work, heat and internal energy, in accordance with the expectation value given by Eq.~\eqref{eq:expectation_notation}. This limit is important for thermodynamic considerations: 
Eq.~\eqref{eq:ent_prod_clausius} has the form of a standard Clausius's statement of the second law, if the temperatures and the chemical potentials of the baths can be considered constant. If the baths are macroscopic, they have an infinite capacity for heat and particles, and hence their macroscopic properties temperatures and chemical potentials can be considered constant. So, for infinitely large baths Eq. \eqref{eq:ent_prod_relative_ent} is indeed a statement of the second law. On the other hand, for finite bath sizes, this constancy assumption regarding the temperatures and chemical potentials does not hold beyond a finite time, thus requiring corrections for a consistent thermodynamic description \cite{Strasberg_2021,Strasberg_tutorial_2021}. 

The above definitions of heat and work seem to suggest that one needs to access the full baths to describe the thermodynamics. However, this is not the case. The heat and work defined above can be written in terms of currents from the baths, which can be obtained without having access to the full microscopic details of the baths. The energy current from the $\alpha$-th bath is
\begin{equation}
    \label{eq:energy_current_alpha}
    J^E_\alpha(t) = -\dv{\ev{\hat{H}_\alpha}}{t} = -i\ev{\comm{\hat{H}_{\text{tot}}(t)}{\hat{H}_\alpha}} = i\ev{\comm{\hat{H}_\alpha}{\hat{H}_{\text{S}\alpha}}},
\end{equation}
while the particle current is defined
\begin{equation}
    \label{eq:particle_current_alpha}
    J^P_\alpha(t) = -\dv{\ev{\hat{N}_\alpha}}{t} = -i\ev{\comm{\hat{H}_{\text{tot}}(t)}{\hat{N}_\alpha}} = i\ev{\comm{\hat{N}_\alpha}{\hat{H}_{\text{S}\alpha}}}.
\end{equation}
Finally, the heat current from the $\alpha$-th bath is thus defined as in terms of the energy and particle currents according to
\begin{equation}
    J^Q_\alpha(t) = J^E_\alpha(t) - \mu_\alpha J^P_\alpha(t).
\end{equation}
Using these definitions, the heat dissipated into the $\alpha$-th bath and the chemical work can be recast explicitly in terms of these currents as
\begin{align}
\label{eq:in_terms_of_currents}
& Q_\alpha(t) = -\int_0^t dt^\prime  J^Q_\alpha(t^\prime), \quad ~W_\text{chem}(t) = \sum_\alpha \mu_\alpha \int_0^t dt^\prime J^P_\alpha(t^\prime), \nonumber \\
& W_\text{ext}(t)= \delta U(t) - \int_0^t dt^\prime J^E_\alpha(t^\prime),
\end{align}
where the last expression follows from the first law.
These expressions, along with Eqs.~\eqref{eq:ent_prod_clausius} and \eqref{eq:internal_energy}, show that the description of average thermodynamics quantities can be described by knowing, as function of time, the energy and the particle currents from the baths, the state of the system and the system-bath coupling energy. 

It is important to mention that the above definitions of thermodynamic quantities, while being completely self-consistent, are not unique, especially at strong coupling between the system and the baths~\cite{Hanggi_strong_couple}. In particular, there are valid criticisms against the definition of $\delta U(t)$ including the system-bath coupling energy, and the identification of von-Neumann entropy as the thermodynamic entropy. However, if the system is finite-dimensional and is driven (either by an external drive, or via chemical potential and temperature biases, or both), the work and heat increase indefinitely, while the change in internal energy $\delta U(t)$ and the change in entropy of the system $\delta S_\text{S}(t)$ remain finite, regardless of their actual definitions. Consequently, $\delta U(t)$ and $\delta S_\text{S}(t)$ become negligible in the expressions of external work [Eq.~\eqref{eq:in_terms_of_currents}] and entropy production [Eq.~\eqref{eq:ent_prod_clausius}] respectively. As a result, in such cases, the long-time thermodynamics can be entirely and uniquely described in terms of the currents from the baths.

So far the formalism we have described holds for arbitrary bath Hamiltonians. Next we focus specifically on the case where each bath is described by an infinite number of fermionic or bosonic modes prepared in a thermal state. 
In this case, the Hamiltonian of the $\alpha$-th bath is given by
\begin{equation}
    \label{eq:hamiltonian_bath_macro}
    \hat{H}_\alpha = \sum_{m=1}^\infty \omega_{m \alpha} \hat{b}_{m \alpha}^\dagger \hat{b}_{m \alpha},
\end{equation}
where $\hat{b}_{m \alpha}$ are fermionic or bosonic annihilation operators for the $m$-th mode of the $\alpha$-th bath. 
We also fix the nature of the system-bath couplings to
\begin{equation}
    \label{eq:couplings_bath_macro}
    \hat{H}_{\text{S} \alpha} = \sum_{m=1}^\infty \qty(\lambda_{m \alpha} \hat{S}^\dagger_\alpha \hat{b}_{m \alpha} + \lambda_{m \alpha}^* \hat{b}_{m \alpha}^\dagger \hat{S}_\alpha),
\end{equation}
where $\hat{S}_\alpha$ is the system operator coupling with the $\alpha$th bath. With this canonical model of thermal baths, it can be shown that the dynamics of the system, the currents from the baths and the system-bath coupling energy are all uniquely determined by the bath spectral functions,
\begin{equation}
    \mathcal{J}_{\alpha}(\omega) = 2\pi \sum_{m=1}^\infty \abs{\lambda_{m \alpha}}^2 \delta\qty(\omega - \omega_{m \alpha}),
\end{equation}
and the Fermi or the Bose distributions, $f_\alpha(\omega) = (e^{\beta_\alpha(\omega - \mu_\alpha)} \pm 1)^{-1}$.
Crucially, these  properties  do not depend on the fine microscopic details of each bath. Any set of baths with the same spectral functions and same initial temperatures and chemical potentials gives rise to the same dynamics of the system, the currents from the baths, and the system-bath coupling energy. 

While the $L_B \to \infty$ limit is crucial for a consistent description of dynamics and thermodynamics of open quantum systems, it makes the calculation of the above thermodynamic quantities intractable. This is additionally complicated by the fact that, without any further approximations, the reduced dynamics of the system is non-Markovian. Only in absence of external driving, if the system is also quadratic in bosonic or fermionic creation and annihilation operators, the long-time results can be obtained without making further approximations using standard analytical techniques like NEGF, Landauer-Buttiker formalism etc. In presence of external driving, even for quadratic cases, all approaches often have to rely on perturbative techniques (for example, \cite{Ludovico_2016,Bruch_2018,RieraCampeny2019,Brandner_2020}).

In absence of exact analytical techniques, arguably, the most direct way to numerically simulate the dynamics is to consider finite, but large enough, baths and carry out the full unitary evolution. A systematic way to do this is given by the chain-mapping of the baths \cite{rc_mapping_QTD_book,Chain_mapping_bosons_fermions,rc_mapping_bosons,rc_mapping_fermions,Prior,Woods2015, Woods2016, Marscherpa2017, rc_mapping_old} (see Appendix~\ref{appendix:chain mapping}). This utilizes the understanding that we can choose a convenient way to microscopically model the baths, such that the bath spectral functions are left invariant, without changing the dynamics or the thermodynamics. However, the size of the baths required to simulate the dynamics grows with time. If the drive is periodic, and a periodic steady state is reached within the time possible to simulate with the largest size baths, the whole dynamics can be inferred. If this is not the case, for example if the drive is not periodic, the long-time behavior cannot be obtained by this technique.

In the following section, we introduce the mesoscopic leads approach, which allows us to obtain completely non-perturbative numerically exact results, and thereby  accurately describe the thermodynamics of externally driven open quantum systems, up to arbitrarily long times.  
We will benchmark these results against the brute force simulation with the chain-mapping technique.

\section{\label{sec:mesoscopic leads}The mesoscopic leads approach}

\begin{figure}
    \centering
    \includegraphics[width=\columnwidth]{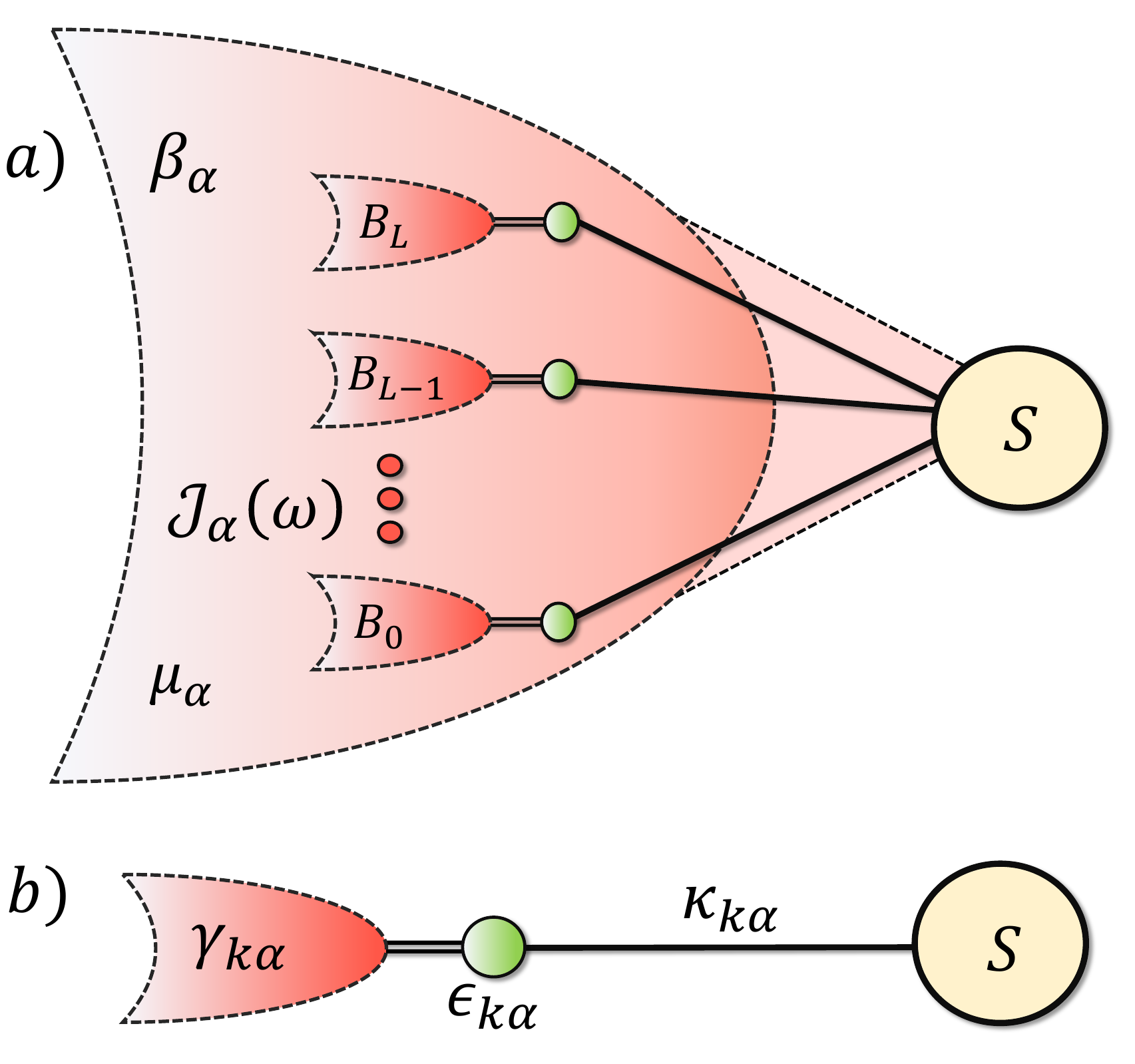}
    \caption{a) The strong coupling between system $S$ and the bath can be can be approximated by $L$ leads which are each independently coupled to independent infinite baths. b) A close up of the one of the leads with energy $\epsilon_{k\alpha}$ and the system, with coupling strength $\kappa_{k\alpha}$. The lead is weakly coupled to the infinite bath with decay rate $\gamma_{k\alpha}$.
    }
    \label{fig:thermleads_hierarchical}
\end{figure}

\subsection{Microscopic derivation}
\label{subsec:mesoscopic_leads_derivation}

We now show how the mesoscopic leads approach ~\cite{Imamoglu_1994,Garraway_1997a,Garraway_1997b,Sanchez_2006,Subotnik_2009,Dzhioev_2011, Ajisaka_2012,Ajisaka_2013,Arrigoni_2013,Dorda_2014,Chen_2014,Zelovich_2014,
Dorda_2015,Hod_2016,Gruss_2016,Schwarz_2016,Dorda_2017,Elenewski_2017,
Gruss_2017,Zelovich_2017,Tamascelli_2018,Lemmer_2018,Oz_2019,Chen_Galperin_2019,
Zwolak_2020,Wojtowicz_2020,Chiang_2020,Brenes2020,Lotem_2020,Fugger_2020,
Wojtowicz_2021,Elenewski_2021} can be microscopically derived in the setting described in the previous section. A schematic of how each bath is microscopically represented in this approach is given in Fig.~\ref{fig:thermleads_hierarchical}(a). This can be done for any given bath spectral function. The first step of this approach is the observation that any continuous bath spectral function can be approximated by discretizing it into $L_\alpha$ (for simplicity, equally spaced) points, $\{\epsilon_k \}$, $k=1$ to $L_\alpha$,  and defining
\begin{equation}
\label{eq:J_eff}
    \mathcal{J}^{\text{eff}}_\alpha(\omega) = \sum_{k=1}^{L_\alpha}\frac{|\kappa_{k\alpha}|^2\gamma_k}{(\omega - \epsilon_{k \alpha})^2 + (\gamma_{k \alpha}/2)^2},
\end{equation}
with 
\begin{align}
\label{eq:thermleads_parameters}
\kappa_{k \alpha} = \sqrt{\frac{\mathcal{J}_{\alpha}(\epsilon_{k\alpha})e_{k\alpha}}{2\pi}},\quad \gamma_{k \alpha} = e_{k \alpha}, \quad e_{k \alpha} = \epsilon_{k+1~\alpha} - \epsilon_{k \alpha}.
\end{align}

With this definition, it can be checked that $\mathcal{J}^{\text{eff}}_\alpha(\omega)$ tends to $\mathcal{J}_{\alpha}(\omega)$ as $L_\alpha$ increases. 
Hence, for a finite but large enough $L_\alpha$ (which corresponds to small enough $e_k$) $\mathcal{J}^{\text{eff}}_\alpha(\omega)$ is a controlled approximation to $\mathcal{J}_{\alpha}(\omega)$.  Eq.~\eqref{eq:J_eff} shows that each bath $\alpha$, with spectral function $\mathcal{J}_{\alpha}(\omega)$, can be decomposed into $L_{\alpha}$ different baths, each with a specific Lorentzian spectral function.

A bath with a Lorentzian spectral density, however, yields the same dynamics as if the system was coupled with strength $\kappa_{k\alpha}$ to an additional fermionic mode  $\hat{a}_{ka}$, with energy $\epsilon_{k\alpha}$, which itself is coupled  to its own bath, with a flat spectral function $\gamma_\alpha$~\cite{Brenes2020}, see Fig.~\ref{fig:thermleads_hierarchical}(b). This can be viewed as a single step of the chain mapping of Appendix~\ref{appendix:chain mapping}, Eq.~\eqref{rc_map}.

We thus arrive at the mesoscopic lead, consisting of $L_\alpha$ modes,
\begin{align}
\hat{H}_{L_\alpha} = \sum_{k=1}^{L_\alpha} \epsilon_{k \alpha} \hat{a}_{k \alpha}^\dagger \hat{a}_{k \alpha},
\end{align}
while the system-bath coupling becomes the system-lead coupling,
\begin{align}
\hat{H}_{S\alpha}=\hat{H}_{SL_\alpha}=\sum_{k=1}^{L_\alpha} \kappa_{k \alpha} \left(\hat{S}_\alpha^\dagger \hat{a}_{k \alpha} + \hat{a}_{k \alpha}^\dagger \hat{S}_\alpha \right).
\end{align}
Now, we define the extended state of the system and lead modes $\hat{\rho}_{\text{ext}}$ and the extended Hamiltonian,
\begin{align}
\hat{H}_\text{ext}(t) = \hat{H}_S(t) + \sum_{\alpha} \left(\hat{H}_{L_\alpha} + \hat{H}_{SL_\alpha}\right),
\end{align}
where $\hat{H}_{L_\alpha}$ is the Hamiltonian of the bath attached to the $\alpha$-th lead mode. The final step is to integrate out the residual baths to obtain an effective equation of motion for the system and the lead modes to the leading order in $e_k$. The crucial point to note in doing so is that $\kappa_{k \alpha} \propto \sqrt{e_{k \alpha}}$ and  $\gamma_{k \alpha} \propto e_{k \alpha}$. So both $H_{SL_\alpha}$ and the coupling between the lead modes and their own residual baths are small in the limit where $\mathcal{J}^{\text{eff}}_\alpha(\omega)$ is a good approximation to $\mathcal{J}_{\alpha}(\omega)$. Using standard techniques \cite{Brenes2020}, this allows us to obtain the following quantum master equation for $\hat{\rho}_{\text{ext}}$,
\begin{equation}
    \label{eq:master_equation}
    \dv{\hat{\rho}_{\text{ext}}}{t} = -i\comm{\hat{H}_\text{ext}(t)}{\hat{\rho}_\text{ext}(t)} + \sum_\alpha \hat{\mathcal{L}}_\alpha\qty{\hat{\rho}_\text{ext}(t)},
\end{equation}
where 
\begin{align}
    \begin{split}
        &\hat{\mathcal{L}}_\alpha\qty{\hat{\rho}_\text{ext}(t)}   \\
        &=\sum_{k=1}^{L_\alpha} \gamma_{k \alpha} e^{\beta_\alpha(\epsilon_{k \alpha} - \mu_\alpha)} f_{k \alpha} \qty[\hat{a}_{k \alpha} \hat{\rho}_\text{ext}(t) \hat{a}_{k \alpha}^\dagger -\frac{1}{2}\acomm{\hat{a}_{k \alpha}^\dagger \hat{a}_{k \alpha}}{\hat{\rho}_\text{ext}(t)}] \\
        &+\sum_{k=1}^{L_\alpha} \gamma_{k\alpha} f_{k \alpha} \qty[\hat{a}_{k \alpha}^\dagger \hat{\rho}_\text{ext}(t) \hat{a}_{k \alpha} -\frac{1}{2}\acomm{\hat{a}_{k \alpha} \hat{a}_{k \alpha}^\dagger}{\hat{\rho}_\text{ext}(t)}],
    \end{split}
\end{align}
with $f_{k \alpha} = \qty[e^{\beta_\alpha(\epsilon_{k \alpha} - \mu_\alpha)}\pm 1]^{-1}$ being the Fermi or the Bose distribution function. This is the central equation of the mesoscopic leads approach. Here, each bath is modelled by a finite set of modes, each mode being damped via a local Lindblad operator. The local Lindblad operator is such that it would take the mode to its thermal state if it was uncoupled from the system. To obtain the correct dynamics at all times, the initial state Eq.~\eqref{eq:initial_state} will also have to be closely approximated. For this, we choose
\begin{align}
\hat{\rho}_\text{ext}(0) = \hat{\rho}_{\text{S}}(0)  \prod_\alpha \frac{e^{-\beta_\alpha(\hat{H}_{L_\alpha} - \mu_\alpha \hat{N}_{L_\alpha})}}{Z_{L_\alpha}},
\end{align}
where $\hat{N}_{L_\alpha}=\sum_{k=1}^{L_\alpha}\hat{a}_{k \alpha}^\dagger \hat{a}_{k \alpha}$ is the total number operator for the lead modes, and $Z_{L_\alpha}$ is the corresponding partition function. Clearly, this initial state becomes a close approximation to Eq.~\eqref{eq:initial_state} for large enough lead modes. The above derivation shows  that by increasing the number of lead modes, the dynamics of the system obtained from the above quantum master equation will converge to the dynamics of the system in the presence of baths with spectral functions $\mathcal{J}_{\alpha}(\omega)$.

There are several salient features of the mesoscopic leads approach that make it advantageous over most other approaches. Although this approach has been used in presence of external driving before \cite{Chen_2014,Oz_2019}, these important features, to our knowledge, have hardly been emphasized and appreciated in previous works. The first important point to note is that the above derivation  holds for arbitrary  types of driving in the system. This fact is quite non-trivial, because, in microscopic derivations of quantum master equations, the dissipative part of the quantum master equation can change depending on the nature of the drive \cite{Schnell_2020,Schnell_2021}. However, in the mesoscopic leads approach, the fact that both $\kappa_{k \alpha}$ and  $\gamma_{k \alpha}$ are small allows us to obtain the same dissipators to the leading order in $e_k$ for all kinds of driving.

The small parameter in the derivation is $e_{k \alpha}$, which reduces on increasing the number of lead modes. This is not a small parameter of the physical set-up we want to model, but rather, a small parameter that controls the error in the numerical simulation. Thus, this approach can give fully non-perturbative results, with controlled errors which can be reduced by increasing the number of lead modes. Furthermore, although the individual coupling between each lead mode and the system is small, overall, the effective spectral function $\mathcal{J}^{\text{eff}}_\alpha(\omega)$ is not small. Rather, it is a good approximation to $\mathcal{J}_{\alpha}(\omega)$. The mesoscopic leads approach can thus treat arbitrary strengths of system-bath coupling.  Although the evolution equation for the extended density matrix $\hat{\rho}_{\rm ext}$ is an additive Lindblad equation, the reduced dynamics of the system (i.e, not including the leads) can be fully non-Markovian and have completely non-additive contributions from the leads.

Finally and perhaps most importantly, the extended set-up of the system and the lead modes is of finite size. Due to the Lindblad damping, it incorporates the effect of infinite number of degrees of freedom of the baths. Consequently, unlike the chain-mapping technique mentioned in the previous section, long time simulation can be done while keeping the number of lead modes finite, provided the spectral function is accurately enough represented. This is a crucial advantage of this approach over the brute force method using chain-mapping.

Along with the above salient features, all quantities required for the description of thermodynamics can be obtained from the mescoscopic leads approach, as we discuss in the next subsection.

\subsection{Thermodynamics from mesoscopic leads approach}
\label{subsec:thermodynamics_mesoscopic_leads}

The mesoscopic leads approach simulates the evolution of $\hat{\rho}_{\text{ext}}$. Since this extended state evolves under a Lindblad equation it is Markovian whereas the original system alone still undergoes non-Markovian evolution. This is known as a Markovian embedding. As such, it provides access to expectation values of operators in the extended Hilbert space of system plus leads, as well as the rate of change of expectation values of such operators. We need to write all the quantities required for the description of thermodynamics in terms of such quantities.

As discussed before, we need to know the state of the system, the system-bath coupling energy, and the particle and the energy currents from the baths as a function of time. The state of the system is directly obtained from the mesoscopic leads approach. The system-bath coupling energy is just the expectation value of the system-lead coupling operator, which is also directly obtained because it is an operator in the extended Hilbert space. For obtaining the currents from the baths, we note that the mesoscopic leads approach consider the following microscopic structure of each bath,
\begin{align}
\hat{H}_\alpha = \hat{H}_{L_\alpha} + \hat{H}_{LE_\alpha} + \hat{H}_{E_\alpha},
\end{align}
where $\hat{H}_{E_\alpha}$ describe the composite Hamiltonian of the residual baths of all the lead modes, and $\hat{H}_{LE_\alpha}$ describe the corresponding composite coupling between the lead modes and their residual baths. The particle current from the $\alpha$th bath, defined in Eq.~\eqref{eq:particle_current_alpha}, can now be seen to be given by
\begin{align}
\label{eq:particle_current_expression}
J^P_\alpha(t) = i\ev{\comm{\hat{N}_{L_\alpha}}{\hat{H}_{SL_\alpha}}},
\end{align}
where $\hat{N}_{L_\alpha}=\sum_{k=1}^{L_\alpha}\hat{a}_{k \alpha}^\dagger \hat{a}_{k \alpha}$ is the total number operator for the lead modes. This is also the expectation value of an operator in the extended Hilbert space, so can be directly obtained from the mesoscopic leads approach. However, for energy current, defined in Eq.~\eqref{eq:energy_current_alpha}, we get
\begin{align}
    \label{eq:energy_current_hierarchical}
J^E_{\alpha}(t) =i\ev{\comm{\hat{H}_{L_\alpha} + \hat{H}_{LE_\alpha}}{\hat{H}_{SL_\alpha}}}.
\end{align}
The fact that $\hat{H}_{SL_\alpha}$ and $\hat{H}_{LE_\alpha}$ does not commute does not let us write the energy current as an expectation value in the extended Hilbert space. However, we note that
\begin{align}
    \begin{split}
        \dv{\ev{\hat{H}_{SL_\alpha}}}{t} &= -i\ev{\comm{\hat{H}_{SL_\alpha}}{\hat{H}_{\text{S}}(t)}} -i \ev{\comm{\hat{H}_{SL_\alpha}}{ \hat{H}_{L_\alpha} +  \hat{H}_{LE_\alpha}}} \\
        &= J^E_{\alpha} - i\ev{\comm{\hat{H}_{SL_\alpha}}{\hat{H}_{\text{S}}(t)}},
    \end{split}.
\end{align}
This gives the energy current from the $\alpha$th bath in terms of expectation values of operators on the extended Hilbert space, and rate of change of such operators,
\begin{align}
J^E_{\alpha}(t) = \dv{\ev{\hat{H}_{SL_\alpha}}}{t}+i\ev{\comm{\hat{H}_{SL_\alpha}}{\hat{H}_{\text{S}}(t)}}.
\end{align}
This can therefore be obtained accurately from the mesoscopic lead approach.  Calculating the above expectation values from the quantum master equation, Eq.~\eqref{eq:master_equation} finally yields
\begin{equation}
\label{eq:energy_current_expression}
    J^E_{\alpha}(t) = i\ev{\comm{\hat{H}_{L_\alpha}}{\hat{H}_{SL_\alpha}}} + \Tr[\hat{H}_{SL_\alpha} \hat{\mathcal{L}}_\alpha\qty{\hat{\rho}_{\text{ext}}}].
\end{equation}
    The second term on the right hand side would naively not be expected.

\begin{figure}
    \centering
    \includegraphics[width=\columnwidth]{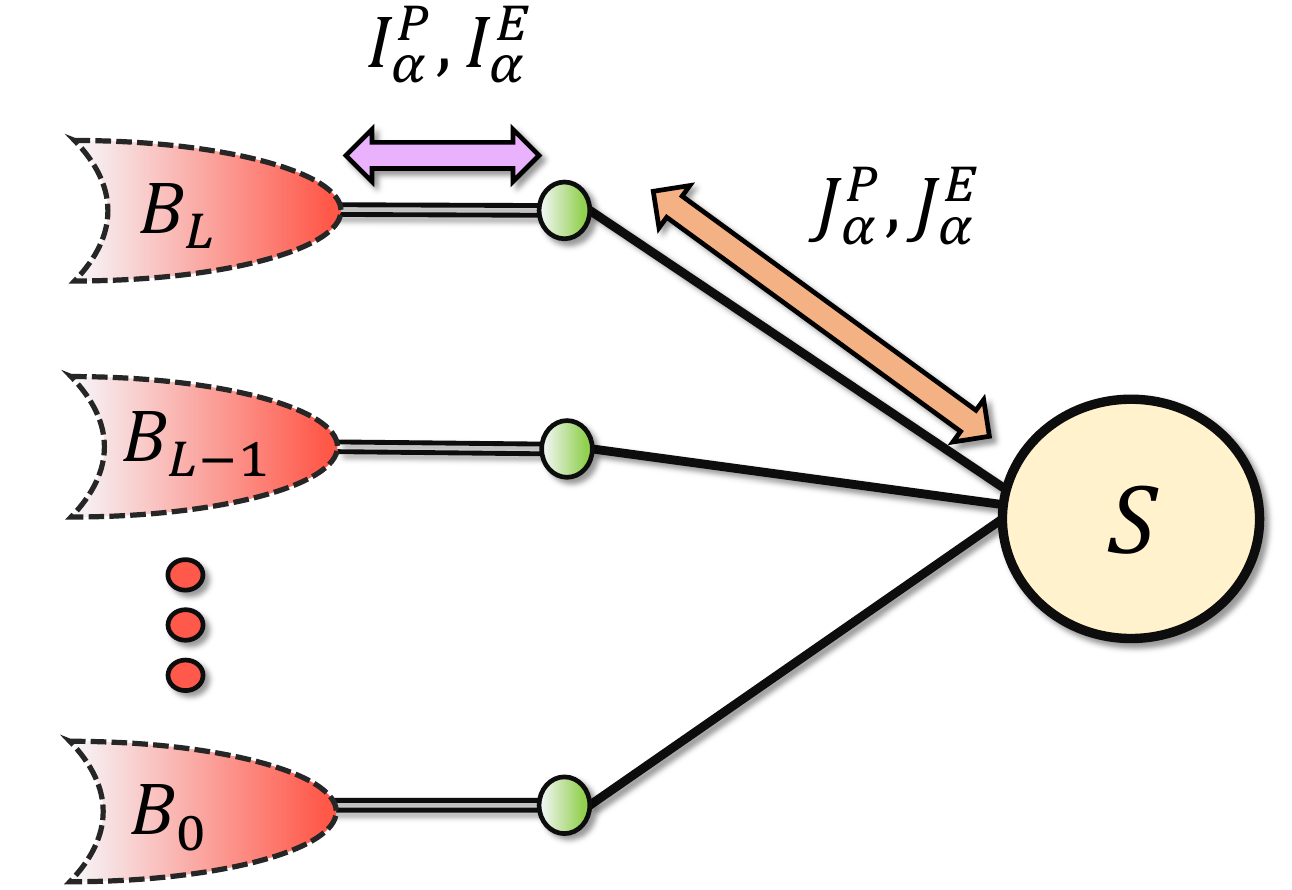}
    \caption{Schematic representation of a system coupled to mesoscopic leads. The arrows represent the two set of currents associated with the left bath, the ``external'' ($I_L^P$ and $I_L^E$) and ``internal'' ($J_L^P$ and $J_L^E$).}
    \label{fig:currents}
\end{figure}

The proper definition of currents is important to describe thermodynamics in the presence of an external drive. Given the quantum master equation, Eq.~\eqref{eq:master_equation}, it is tempting to define the particle and the energy currents as

\begin{align}
    \label{eq:external_currents}
    & I^P_\alpha(t) = \Tr(\hat{N}_{L_\alpha}\hat{\mathcal{L}}_\alpha \qty[\hat{\rho}_\text{ext}(t)]),\quad  ~I^E_\alpha(t) = \Tr(\hat{H}_\text{ext}\hat{\mathcal{L}}_\alpha\qty[\hat{\rho}_\text{ext}(t)]).
\end{align}     

However, these describe currents from the residual baths into the mesoscopic leads. They do not correspond to the current from the baths into the system defined in Eqs. \eqref{eq:particle_current_alpha}, \eqref{eq:energy_current_alpha}, see Fig.~\ref{fig:currents}. For a driven system, $I^P_\alpha(t) \neq J^P_\alpha(t)$, $I^E_\alpha(t) \neq J^E_\alpha(t)$. In the absence of drive, and at the NESS, due to continuity equations, these two different definitions will agree. For similar reasons, with periodic drives, when a periodic NESS (limit cycle) is reached, the time-period average of these two different currents become the same. However, instantaneously they will still be different. The currents from the baths, Eqs. \eqref{eq:particle_current_alpha}, \eqref{eq:energy_current_alpha}, are independent of the microscopic modelling of the baths. If instead of the mesoscopic leads, a different microscopic modelling of the baths giving the same spectral functions is used, like the chain-mapping, the currents would be exactly the same. However, the currents $I^P_\alpha(t)$, $I^E_\alpha(t)$ depend on the particular microscopic modelling of the baths as mesoscopic leads, and may not be replicated in a different microscopic model of the baths, like the chain-mapping.

From the above discussion, it is clear that all quantities required for the description of thermodynamics can be obtained accurately from the mesoscopic leads approach. A crucial point to note is that, throughout the entire discussion till now, we have left the system Hamiltonian and the system operators coupling to the baths completely arbitrary. The entire discussion therefore holds for arbitrary system Hamiltonians, with arbitrary driving.  However, it is non-trivial to simulate the dynamics of $\hat{\rho}_{\text{ext}}(t)$. Simulation in the presence of many-body interactions (i.e, higher than quadratic terms) require  tensor network techniques~\cite{Wojtowicz_2020,Lotem_2020,Brenes2020,Fugger_2020}. While this may be numerically expensive, it is still feasible for a range of parameters where many other techniques fail. And wherever possible, it gives numerically exact non-perturbative results. This has been shown for quantum transport and thermodynamics in absence of any explicit time-dependent drive~\cite{Brenes2020}.

In this work, however, we will  not consider many-body interactions. In absence of many-body interactions, as will be shown below, numerically exact results can be obtained quite elegantly by simulating a Lyapunov equation. For driven systems, such exact results are often not amenable to other numerical or analytical techniques (except for brute force numerics via chain mapping).


\subsection{\label{subsec:lyapunov_eq}The Lyapunov equation for non-interacting systems}

Consider a system Hamiltonian of the form
\begin{equation}
    \hat{H}_\text{S}(t) = \sum_{\ell, m=1}^{L_S}[\mathbf{H_\text{S}}(t)]_{\ell m}\hat{c}^\dagger_\ell \hat{c}_m,
\end{equation}
where $\mathbf{H_\text{S}}(t)$ is a $L_S \times L_S$ Hermitian matrix, sometimes called single-particle Hamiltonian, with time $t$ as a parameter. This describes a number conserving non-interacting fermionic system of $L_S$ sites, in a lattice of arbitrary dimension and geometry with an arbitrary external type of external drive. Consider this system to be coupled via number conserving bilinear coupling with baths at various sites.  In the mesoscopic leads approach, the system-lead coupling for the lead describing the bath attached at the $\alpha$th site is
\begin{align}
\hat{H}_{SL_\alpha}=\sum_{k=1}^{L_\alpha} \kappa_{k \alpha} \left(\hat{c}_\alpha^\dagger \hat{a}_{k \alpha} + \hat{a}_{k \alpha}^\dagger \hat{c}_\alpha \right).
\end{align}
The extended set-up of system and  leads can now be written in the form
\begin{align}
\hat{H}_{\text{ext}}(t) = \sum_{\ell, m=1}^{L_S+\sum_\alpha L_\alpha}[\mathbf{H_\text{ext}}(t)]_{\ell m}\hat{d}^\dagger_\ell \hat{d}_m,
\end{align}
where $\hat{d}_m$ is a fermionic annihilation operator of the either a system site or a lead site. For such cases, if the initial state of the system is Gaussian, the dynamics of the whole set-up remains Gaussian at all times. All properties of the extended system can be obtained by calculating the correlation matrix, also called the single-particle density matrix, $\mathbf{C}(t)$, with elements~\cite{Landi_Poletti_Schaller_2021}
\begin{align}
\mathbf{C}_{pq}(t)=\Tr[\hat{d}_q^\dagger \hat{d}_p \hat{\rho}_{\text{ext}}(t)].
\end{align}
Since $\hat{\rho}_{\text{ext}}(t)$ is Gaussian at all times, it can be obtained exactly from $\mathbf{C}_{pq}(t)$. The crucial simplification is that one can directly write down the equation of motion for $\mathbf{C}_{pq}(t)$ from Eq.~\eqref{eq:master_equation}, in the form of a Lyapunov equation, as we discuss below. For simplicity and relevance to later examples, in the following, we consider two baths, but the discussion can be straightforwardly generalized to more than two baths.

For two baths, the matrices  $\mathbf{H_\text{ext}}(t)$ and $\mathbf{C}(t)$ can be written in the block form
\begin{align}
&\mathbf{H_{\text{ext}}}(t)=\left[
\begin{array}{ccc}
\mathbf{H}_{\text{S}}(t) & \mathbf{H}_{SL_1} & \mathbf{H}_{SL_2} \\
\mathbf{H}_{SL_1}^\dagger & \mathbf{H}_{L_1} & \mathbf{0} \\
\mathbf{H}_{SL_2}^\dagger & \mathbf{0} & \mathbf{H}_{L_2}
\end{array}
\right], \nonumber \\
&\mathbf{C}(t)=\left[
\begin{array}{ccc}
\mathbf{C}_{\text{S}}(t) & \mathbf{C}_{SL_1}(t) & \mathbf{C}_{SL_2}(t) \\
\mathbf{C}_{SL_1}^\dagger(t) & \mathbf{C}_{L_1}(t) & \mathbf{C}_{L_1 L_2}(t) \\
\mathbf{C}_{SL_2}^\dagger(t) & \mathbf{C}_{L_1 L_2}^\dagger(t) & \mathbf{C}_{L_2}(t)
\end{array}
\right].
\end{align}
Here, the $L_S \times L_\alpha$ matrix $\mathbf{H}_{SL_\alpha}$  gives the coupling between the system and the lead representing the $\alpha$th bath, 
and $\mathbf{H}_{B_\alpha}=\text{diag}\{\epsilon_{k\alpha}\}$  gives the Hamiltonian of the $\alpha$th lead. Similarly, the $L_S \times L_S$ dimensional matrix $\mathbf{C}_S(t)$ gives the system correlation matrix, the $L_S \times L_\alpha$ dimensional matrix   $\mathbf{C}_{SL_\alpha}(t)$  gives the correlations between the system and the $\alpha$th lead, $\mathbf{C}_{L_\alpha}(t)$ gives the correlation matrix corresponding to the $\alpha$th lead, and $\mathbf{C}_{L_1 L_2}(t)$ gives the correlations between the two leads.
With the correlation matrix written in this form, its equation of motion, obtained from Eq.~\eqref{eq:master_equation}, is given by the following continuous-time differential Lyapunov equation
\begin{equation}\label{eq:lyapunov}
    \dv{\mathbf{C}}{t} = -\left(\mathbf{W}(t) \mathbf{C}(t) + \mathbf{C}(t) \mathbf{W}^\dagger(t)\right) + \mathbf{F},
\end{equation}
with $\mathbf{W}(t)=i\mathbf{H_\text{ext}}(t)+\mathbf{\Upsilon}/2$, where
\begin{align}
\mathbf{\Upsilon}=\left[
\begin{array}{ccc}
\mathbf{0} & \mathbf{0} & \mathbf{0} \\
\mathbf{0} & \mathbf{\Upsilon}_{1} & \mathbf{0} \\
\mathbf{0} & \mathbf{0} & \mathbf{\Upsilon}_{2}
\end{array}
\right],~~
\mathbf{F}=\left[
\begin{array}{ccc}
\mathbf{0} & \mathbf{0} & \mathbf{0} \\
\mathbf{0} & \mathbf{F}_{1} & \mathbf{0} \\
\mathbf{0} & \mathbf{0} & \mathbf{F}_{2}
\end{array}
\right]
\end{align}
and $\mathbf{\Upsilon}_{\alpha}=\text{diag}\{\gamma_{k\alpha}\}$, $\mathbf{F}_{\alpha}=\text{diag}\{\gamma_{k\alpha} f_{k\alpha}\}$. The dynamics can be obtained by integrating this matrix equation numerically, say via Runge-Kutta methods. The size of the matrices required scales linearly with total number of sites in the system and the leads, and not exponentially, as it would have been if we were to directly consider the evolution of $\hat{\rho}_{\text{ext}}(t)$. All quantities required for thermodynamics can also be obtained by knowing $\mathbf{C}(t)$. Moreover, this is possible for arbitrary types of driving, which is beyond the reach of most other analytical or numerical techniques.
Further simplification is possible if the drive is periodic, as we show in the next subsection.

\subsection{\label{subsec:floquet}Floquet solution of the periodically-driven Lyapunov equation}

The dynamics of Eq.~\eqref{eq:lyapunov}, in the presence of a periodic drive, will be characterized by a transient, followed by a limit cycle where $\mathbf{C}(t)$ becomes time-periodic. 
If one is interested only in the limit cycle, naively this would require solving Eq.~\eqref{eq:lyapunov} over many periods, which may be expensive. 
Instead, one can employ the following method
For simplicity, we assume the drive takes the form 
$\mathbf{W}(t) = \mathbf{W}_0 + \mathbf{W}_1 \cos(\omega t)$. The generalisation to multiple harmonics is straightforward. 
We then attempt a solution of the form $\mathbf{C}(t) = \sum_{n=-\infty}^\infty \mathbf{C}_n e^{i n \omega t}$. Hermiticity of $\mathbf{C}$ implies that $\mathbf{C}_n^\dagger = \mathbf{C}_{-n}$. 
Plugging this in Eq.~\eqref{eq:lyapunov} yields a set of recursive algebraic equations 
\begin{align}\label{eq:floquet_lyapunov}
    in \omega \mathbf{C}_n + \mathbf{W}_0 \mathbf{C}_n + \mathbf{C}_n \mathbf{W}_0^\dagger = \mathbf{F}\delta_{n,0} &- \frac{1}{2} \mathbf{W}_1 (\mathbf{C}_{n-1} + \mathbf{C}_{n+1}) 
    \\[0.2cm]
    &- \frac{1}{2} (\mathbf{C}_{n-1} + \mathbf{C}_{n+1})\mathbf{W}_1^\dagger.
    \nonumber
\end{align}
This  can now be solved iteratively, as in the Gauss-Seidel method~\cite{Gauss_siedel}.
First, one takes $\mathbf{C}_{n\neq 0} = 0$  and solve Eq.~\eqref{eq:floquet_lyapunov} for $\mathbf{C}_0$. 
Then the result is used to set up 3 new equations for $\mathbf{C}_{-1},\mathbf{C}_0,\mathbf{C}_1$, which in turn is used to set up 5 equations for 
$\mathbf{C}_{-2},\mathbf{C}_{-1},\mathbf{C}_0,\mathbf{C}_1,\mathbf{C}_{2}$, and so on.

The solution is greatly facilitated when $\mathbf{W}_0$ is diagonalizable (which is almost always the case); that is, $\mathbf{W}_0 = \mathbf{S} \mathbf{\Lambda} \mathbf{S}^{-1}$, where $\mathbf{\Lambda}$ is a diagonal matrix containing the eigenvalues (which are generally complex). 
Define $\tilde{\mathbf{C}}_n = \mathbf{S}^{-1} \mathbf{C}_n (\mathbf{S}^{-1})^\dagger$ and similarly for $\tilde{\mathbf{F}}$ and $\tilde{\mathbf{W}}_1$. 
Then Eq.~\eqref{eq:floquet_lyapunov} can be rewritten element-wise as 
\begin{align}\label{eq:floquet_lyapunov_eigen}
    [\tilde{\mathbf{C}}_n]_{i,j}  = \frac{1}{i n \omega + \Lambda_i + \Lambda_j^*} \Big[ \delta_{n,0}\tilde{\mathbf{F}}
    &- \frac{1}{2}  \tilde{\mathbf{W}}_1 (\tilde{\mathbf{C}}_{n-1} + \tilde{\mathbf{C}}_{n+1}) \\[0.2cm]
    &- \frac{1}{2} 
    (\tilde{\mathbf{C}}_{n-1} + \tilde{\mathbf{C}}_{n+1})\tilde{\mathbf{W}}_1 \Big]_{i,j}.
\nonumber    
\end{align}
The reason why this is advantageous is because the largest overhead in Eq.~\eqref{eq:floquet_lyapunov} is the linear system of equations associated with the fact that $\mathbf{W}_0$ is not diagonal.
In Eq.~\eqref{eq:floquet_lyapunov_eigen}, the only computationally expensive part is the diagonalization of $\mathbf{W}_0$, which only has to be performed once. Afterwards, all operations involve only simple matrix multiplications. 
After convergence, one recovers $\mathbf{C}_n = \mathbf{S} \tilde{\mathbf{C}}_n \mathbf{S}^\dagger$ and use this to construct  $\mathbf{C}(t) = \sum_{n=-\infty}^\infty \mathbf{C}_n e^{i n \omega t}$. This yields the solution within the limit cycle; that is, valid through an entire period of oscillation. We now test out our technique on a number of models that highlight the power and versatility of the methodology to explore thermodynamics beyond the reach of most other approaches. 

\begin{figure*}[t]
    \centering
    \includegraphics{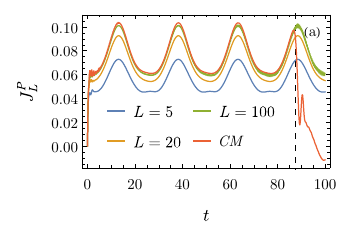}
    \includegraphics{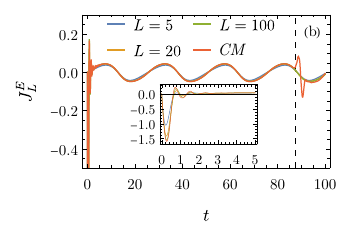}
    \includegraphics{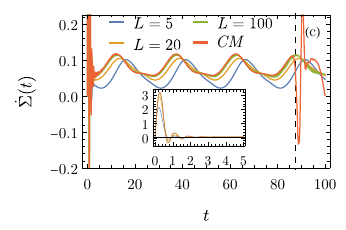}
    \caption{(a) Particle current as a function of time in the driven-resonant model coupled to two baths for increasing number of modes $L$ in each lead, where both leads have the same number of modes. The red curve shows the results obtained the chain-mapping (CM) simulation with $L=100$, which is exact up to a finite time, indicated with a dashed line. (b) Same as (a), but for the energy current. The inset shows a zoomed vision of the short-time scale. (c) Same as (a) and (b), but for the entropy production rate. In these calculations we used $\Gamma=1$, $W=8$, $W^*=4$, $T_L=T_R=1$, $\mu_L=-\mu_R=0.5$, $A=1$, $\omega = 0.25$, $p_0=0.5$ and $L_\text{lin}/L_\text{log}=0.1$.}
    \label{fig:resonant_level_currents}
\end{figure*}

\begin{figure}
    \centering
    \includegraphics{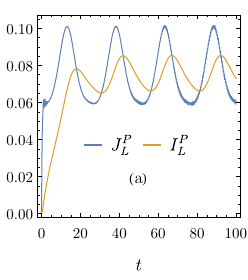}
    \includegraphics{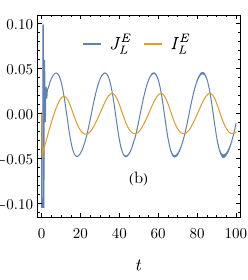}
    \caption{(a) Comparison of the particle current from the left bath in the two-terminal driven resonant level with the currents from the residual baths into the mesoscopic leads. (b) Same, but for the energy currents. In these calculations, both leads had $L=100$ modes and the other parameters were $\Gamma=1$, $W=8$, $W^*=4$, $T_L=T_R=1$, $\mu_L=-\mu_R=0.5$, $A=1$, $\omega = 0.25$, $p_0=0.5$ and $L_\text{lin}/L_\text{log}=0.1$.}
    \label{fig:external_vs_internal}
\end{figure}

\begin{figure}[t]
    \centering
    \includegraphics{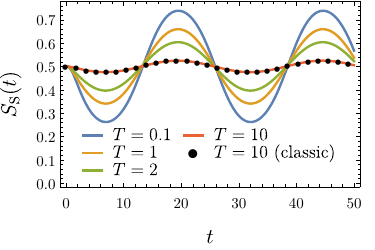}
    \caption{Convergence of the entropy of the dot in the driven resonant-level model with two terminals using the mesoscopic lead approach (solid lines) towards the value predicted by the Pauli master equation [Eq~\eqref{eq:pauli_master_eq}] (black dots) upon increasing the value of temperature, where the temperatures of both baths are set to the same value. The amplitude of the driving is $A=1$ and the driving frequency is $\omega=25$. In the mesoscopic lead simulations the other parameters were $W=8$, $W^*=4$, $L=100$, $L_\text{log}/L_\text{lin}=0.1$,  $\Gamma=1$ and $p_0=0.5$.}
    \label{fig:high_temperature_limit}
\end{figure}

\section{\label{sec:resonant_level}Thermodynamics of a driven resonant-level}

As a first example, we apply the formalism to study the thermodynamics of the driven resonant-level model \cite{Esposito_2015_1,Esposito_2015_2,Ludovico_2016,Ludovico_2016_2,
Bruch_2018,Haughian_2018,Oz_2019}. This provides a simple model in which we can benchmark our results against other methodologies employed in the literature. 
The system is composed of a single dot with an externally controlled, time-dependent energy $\epsilon(t)$. The Hamiltonian is
\begin{equation}
    \label{eq:resonant_level_hamiltonian}
    \hat{H}_\text{S}(t) = \epsilon(t)\hat{c}^\dagger \hat{c},
\end{equation}
where $\hat{c}^\dagger$ and $\hat{c}$ are the fermionic creator and annihilator operators of the system. 
We focus on the two-terminal case, in which the dot is coupled to two baths, labelled by L and R. Our framework can, in principle, handle the case in which each bath is described by an arbitrary, structured spectral density. For simplicity, however, we consider the case in which both baths are described by the same flat spectral density, given by
\begin{equation}
\label{eq:flat_spectral_density}
\mathcal{J}(\omega) =
    \begin{cases}
        \Gamma, & \omega \in \qty[-W, W]\\
        0, & \text{otherwise},
    \end{cases}
\end{equation}
where $\Gamma$ is the system-bath coupling strength and $W$ is a hard cut-off.

Within the mesoscopic leads approach, each bath is described by a damped lead with $L$ modes. 
Fixing a set of energies $\{\epsilon_k\}$ that appropriately sample the spectral density, the couplings $\{\kappa_{k\alpha}\}$ and the damping rates $\{\gamma_{k\alpha}\}$ are fixed by Eq.~\eqref{eq:thermleads_parameters}. Here, we follow Refs.~\cite{Schwarz_2016,Weichselbaum_2009} and choose the energies $\{\epsilon_k\}$ via linear-logarithmic discretization. In this procedure, $L_\text{lin}$ energies are chosen to be linearly spaced in some frequency window $[-W^*, W^*]$, then $L_\text{log}$ energies are chosen to be logarithmically spaced in each of the intervals $[-W, W^*]$ and $[W^*, W]$, so that the total number of modes is $L_\text{lin} + 2L_\text{log}$. For a careful discussion on the choice of discretization procedure, see Ref.~\cite{Elenewski2021}. Throughout all the examples in this section, we fix $\Gamma=1$ and use this to set the energy scale. Furthermore, we fix $W=8$ and $W^*=4$.

We take the energy of the dot to be driven by a sine-protocol of the form
\begin{equation}
    \epsilon(t) = A \sin(\omega t).
\end{equation}
We study the full time-evolution of the extended system by numerically integrating the Lyapunov equation~\eqref{eq:lyapunov}. In Fig.~\ref{fig:resonant_level_currents} (a) and (b), we show the particle and energy current, respectively, computed with $A=1$ and $\omega = 0.25$. We set the baths to have equal temperature $T_\text{L} = T_\text{R} = 1$, but opposite chemical potentials $\mu_\text{L} = -\mu_\text{R} = 0.5$, and the initial population of the dot to be $p_0=0.5$. We take both leads to have the same number of modes $L$, and show the results for increasing values of $L$.  
For comparison, we also present  results obtained via brute-force simulation with chain mapping (Appendix~\ref{appendix:chain mapping}), which is numerically exact only up to a finite time, determined by the size of the chain taken, indicated with a vertical dashed lines. It is completely clear that with increase in number of lead modes, the results converge to that obtained by chain-mapping method up to the time it remains valid. In Fig.~\ref{fig:resonant_level_currents} (b), the inset shows a zoom of the short time-scale. Remarkably, even the short-time dynamics is well described.  Note that, the harmonically driven resonant level with constant bath spectral functions can be exactly solved \cite{Jauho_book}. This would yield exactly the same results as obtained from the chain-mapping procedure up to a finite time proportional to bath size.

It is worth re-stating that in the mesoscopic-leads formalism one has full access to the state of the system at all times and for all driving frequencies. This is in contrast to most other formalisms, like NEGF, Landauer-B\"uttiker formalism, where it is difficult to calculate dynamics at all times and for all driving frequencies. This means, for instance, that we can compute the entropy production rate at all times, and not only the average over a cycle. In the case of the resonant-level, since the correlation matrix reduces to only a number, the entropy of the system is given by $S_\text{S}(t) = -p_t\log p_t - (1-p_t)\log(1-p_t)$, where $p_t = \langle c^\dagger c \rangle$ is the population of the dot at time $t$. Therefore, by simply differentiating this function with respect to time, we can compute the instantaneous entropy production rate [c.f. Eq.~\eqref{eq:ent_prod_clausius}] as $\dot\Sigma(t) = \dot{S}_\text{S}(t) - \sum_\alpha \beta_\alpha J_\alpha^Q$. This is shown in Fig.~\ref{fig:resonant_level_currents} (c), where convergence to chain-mapping results up to a finite time is also shown. 

The convergence of results obtained from mesoscopic leads to those obtained by chain-mapping also demonstrates the fact these results are independent of the particular microscopic modelling of the baths. All microscopic models of the baths which give the same spectral functions lead to same dynamics and thermodynamics at all times. Using the proper definition of currents, given in Eqs.\eqref{eq:particle_current_expression} and \eqref{eq:energy_current_expression}, is crucial for this. If, instead, we used Eq.~\eqref{eq:external_currents}, which is tempting to use given the Lindblad description of mesoscopic leads, the converged result from mesoscopic leads approach would be different, as shown in Fig.~\ref{fig:external_vs_internal}. Therefore, they do not agree with results from the chain-mapping approach. The expressions in Eq.~\eqref{eq:external_currents} give currents from the residual baths to the lead modes (see Fig.~\ref{fig:currents}) which depend on the particular microscopic modelling on the baths.


Next, as an important check, we compare the quantities computed for the driven resonant-level model against their classical counterpart in the high-temperature. 
We consider the same setup as before, with exactly the same parameters except for the temperature of the two baths. In the limit $T \gg \Gamma$, the population of the dot is well described by a classical Pauli master equation of the form
\begin{equation}
    \label{eq:pauli_master_eq}
    \dv{p_t}{t} = \sum_{\alpha=\text{L,R}}\big[ -\Gamma(1 - f_\alpha(t)) p_t + \Gamma f_\alpha(t)(1 - p_t)\big],
\end{equation}
with a time-dependent Fermi distribution $f_{\alpha}(t) = [e^{(\epsilon(t) - \mu_\alpha)/T}+1]^{-1}$. 
In Fig.~\ref{fig:high_temperature_limit}, we show the comparison of the entropy of the dot computed using our approach for increasing values of $T$ against the results obtained by numerically integrating the classical master equation [Eq.~\eqref{eq:pauli_master_eq}] for high temperature. The plots clearly show the convergence of the results with increasing temperature, providing another benchmark for the consistency of our approach. This simply demonstrates that the mesoscopic leads approach recovers the high-temperature, classical limit. At lower temperatures and weak-coupling, one could also benchmark the mesoscopic leads approach against a Floquet-Born-Markov approach \cite{KOHLER2005379, PhysRevA.99.063828}, which has a wider regime of validity. In fact, the formalism in Sec.~\ref{subsec:floquet} is a version of Floquet-Born-Markov approach, but applied to the extended set-up of system and the leads. However, since we have already benchmarked against exact results, we proceed to a more non-trivial system in the next section.

\begin{figure}
    \centering
    \includegraphics[width=\columnwidth]{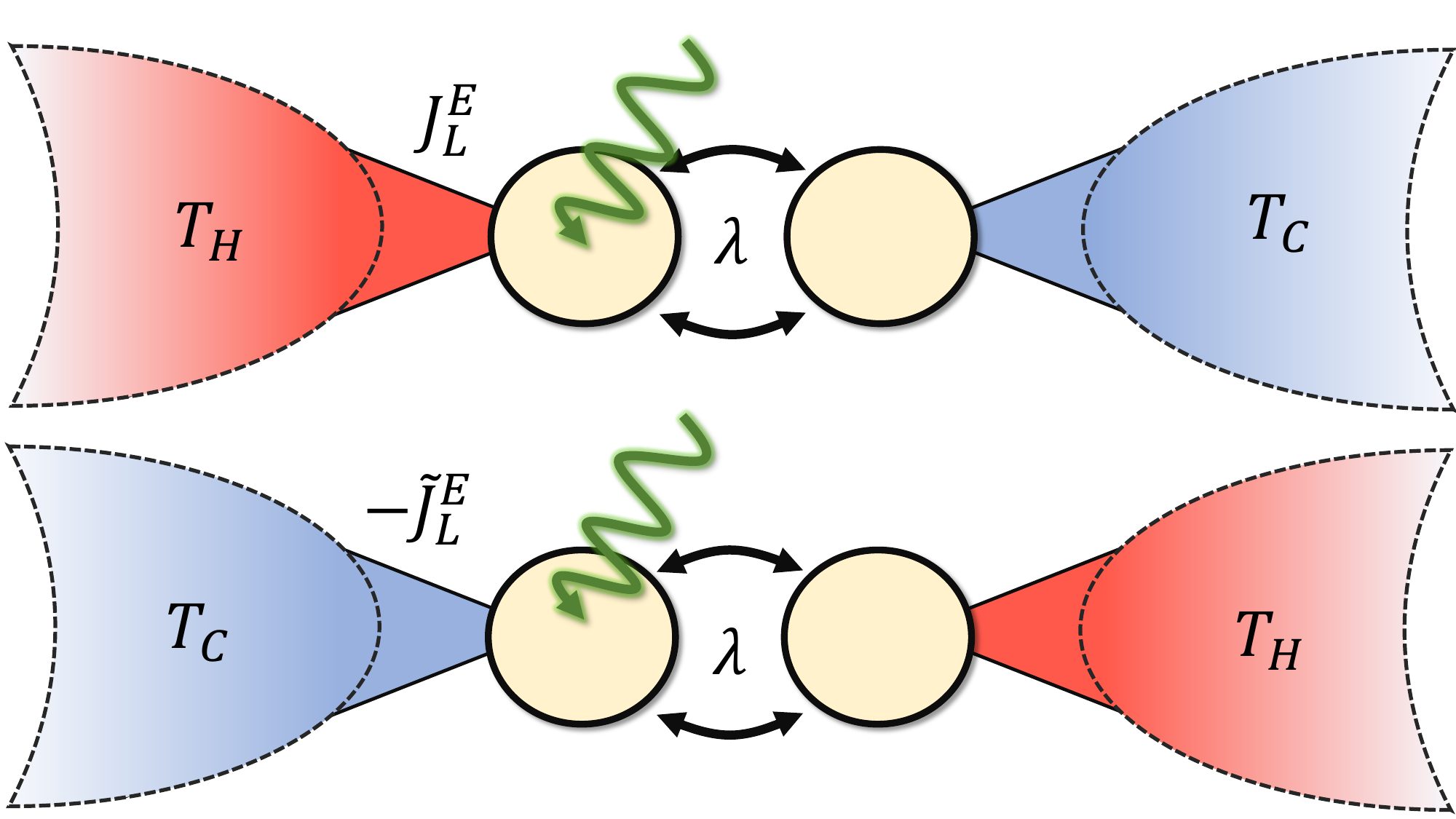}
    \caption{Schematic of the rectification model. Two double dots are coupled to one another with coupling strength $\lambda$. Here the left dot is driven by a time dependent, oscillatory force at a frequency $\omega$. We consider the difference in the particle current when the hot and cold baths are swapped.}
    \label{fig:rectification}
\end{figure}

\section{\label{sec:rectification}Entropy cost of energy rectification in a driven non-interacting system}

Motivated by Ref.~\cite{RieraCampeny2019}, we study heat rectification in a periodically driven non-interacting system. 
There, the authors studied the problem for bosonic degrees of freedom. Here, we consider instead a chain of two fermionic modes, in which only the first mode is driven. A similar problem was also considered in Ref.~\cite{Kohler_2004,Chen_2014}, but focusing only on particle currents. With our framework, we can now also extend this to energy current, and hence heat rectification~\cite{Li_2012}.

The Hamiltonian of the system is given by
\begin{equation}
    \hat{H}_\text{S}(t) = \epsilon_1(t)\hat{c}^\dagger_1 \hat{c}_1 + \epsilon_2\hat{c}^\dagger_2 \hat{c}_2+ \lambda \qty(\hat{c}^\dagger_1 \hat{c}_2 + \hat{c}^\dagger_2 \hat{c}_1),
\end{equation}
where $\lambda$ is the coupling strength between the two modes, $\epsilon_2$ is the energy of the second and $\epsilon_1(t)$ is the energy of the first dot, which is controlled by periodic protocol of the form $\epsilon_1(t) = \epsilon_1(0) + A\sin(\omega t)$. In our simulations, we always fixed $\epsilon_1(0) = \epsilon_2 = 0$. Additionally, each mode is coupled to its own bath. By simplicity, we take both baths to be described by the same flat spectral density, given by Eq.~\eqref{eq:flat_spectral_density}. Again, we fix $\Gamma=1$ for both baths, and use this parameter to set the energy scales. We also choose the chemical potentials of both leads to be zero, so that the heat current matches the energy current.

In order to study heat rectification, two configurations of the setup are considered. In the forward configuration, we set $T_\text{L} = T_\text{hot}$ and $T_\text{R} = T_\text{cold}$, where $T_\text{hot} > T_\text{cold}$. In the reverse setup, the temperatures are swapped, so $T_\text{L} = T_\text{cold}$ and $T_\text{R}=T_\text{hot}$ (see Fig.~\ref{fig:rectification}).
Since only the first dot is driven, the system is not symmetric with respect to left-right inversion. As a consequence, the currents in the two scenarios will  generally be different.

\begin{figure}
    \centering
    \includegraphics{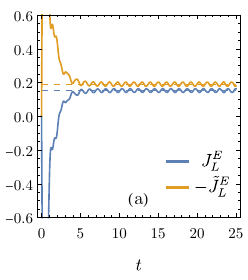}
    \includegraphics{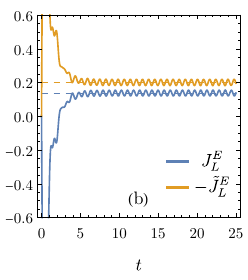}\\
    \includegraphics{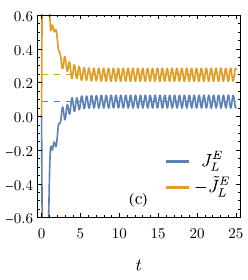}
    \includegraphics{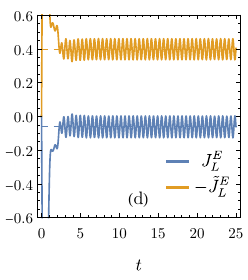}
    \caption{Energy current as a function of time in the forward configuration (blue curve) and opposite of the energy current as a function of time in the reverse configuration (orange curve) for different values of driving frequency: (a) $\omega=3$ (b) $\omega=4$ (c) $\omega=5$ (d) $\omega=6$. The dashed lines indicate the average current in the limit cycle. The coupling strength between the two modes if fixed to $\lambda=3$ and the other parameters are $T_\text{hot}=2$, $T_\text{cold}=1$, $A=1$, $\Gamma=1$, $\mu_\text{L}=\mu_\text{R}=0$, $W=8$, $W^*=4$ and $L=100$ with $L_\text{lin}/L_\text{log} = 0.1$}
    \label{fig:energy_current_rectification}
\end{figure}

\begin{figure}
    \centering
    \includegraphics{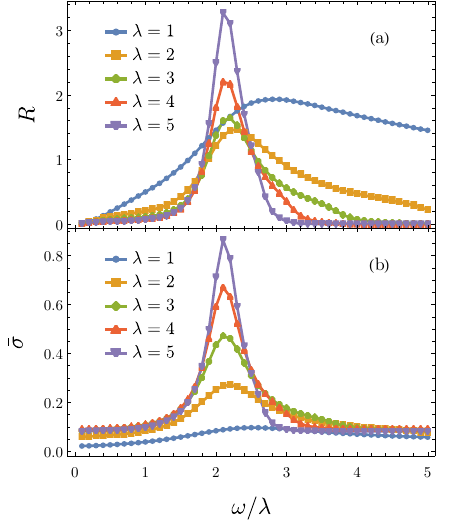}
    \caption{(a) Rectification coefficient as a function of the driving frequency divided by the coupling strength $\lambda$ between the two modes $\lambda$, for increasing values of $\lambda$. (b) Cycle-averaged entropy production rate as a function of the driving frequency divided by the coupling strength between the two modes $\lambda$, for increasing values of $\lambda$. The parameters used are $T_\text{hot}=2$, $T_\text{cold}=1$, $A=1$, $\Gamma=1$, $\mu_\text{L}=\mu_\text{R}=0$, $W=8$, $W^*=4$ and $L=100$ with $L_\text{lin}/L_\text{log} = 0.1$}
    \label{fig:rectification_coef}
\end{figure}

Since the system is time-dependent, one must be careful with the notion of rectification, since the  heat flowing through the system is ill-defined. Due to the presence of the driving, the magnitude of the time period averaged energy currents associated with the left and right bath are not the same, even in the long-time limit. In fact, they differ precisely by the time period averaged power associated with the driving. For this reason, we shall focus only on the energy current flowing towards the left mode, which is the one being driven. As a particular example, let us choose $\lambda=3$. 
We use the method described in Sec.~\ref{subsec:lyapunov_eq} to plot the energy currents in the forward and reversed configuration as a function of time for increasing values of $\omega$, which is shown in Fig.~\ref{fig:energy_current_rectification}. In our simulations, the amplitude of the driving is $A$ and the temperatures are $T_\text{hot}=2$ and $T_\text{cold}=1$. We fixed the number of modes in both leads to be $L=100$, which was checked to be enough to ensure proper convergence of the simulations. We denote by $J^E_L$ the energy current from the left bath in the forward configuration, while by $-\tilde{J}^E_L$ energy current into the left bath in the backward configuration (see Fig.~\ref{fig:rectification}). Both currents eventually go into a limit cycle show periodic oscillations with the same frequency as the drive. The horizontal dashed lines correspond to the cycle-average values in the limit cycle, computed using the method of Sec.~\ref{subsec:floquet}.  The plots clearly show that the magnitude of the forward and backward currents are not the same in the limit cycle. Instead, the gap between them depends on the driving frequency. We have found that analogous rectification does not occur in the particle current for our chosen set-up.

Another interesting point to note is that, in Fig.~\ref{fig:energy_current_rectification}(d), by our sign  conventions, the cycle-averaged heat flows into the left bath in both forward and backward configurations. In forward configuration the left bath is hot (see Fig.~\ref{fig:rectification}). So, it shows that, due to the relatively high frequency of the drive, the hot bath is getting heated in this configuration.

In order to better study the frequency dependence of rectification, we define the rectification coefficient as
\begin{equation}
    R = \frac{\abs{\overline{J^E_L} + \overline{\tilde{J}^E_L}}}{\abs{\overline{J^E_L} - \overline{\tilde{J}^E_L}}},
\end{equation}
where $\overline{J^E_L}$ ($\overline{\tilde{J}^E_L}$) is the cycle-averaged energy current in the forward (backward) configuration in the long-time limit. In the case of a perfect diode, i.e., where there is some current flowing in one configuration but none in the other, we have that $R=1$. If there is no rectification, i.e., both currents have the same magnitude but opposite signs, then $R=0$. Interestingly, this quantity, which is analogous to that used to characterize rectification in absence of external driving, is not upper bounded by $1$. Due to external driving, it can happen that the direction of cycle-averaged heat current is the same in both configurations, as we saw in Fig.~\ref{fig:energy_current_rectification}(d). In such cases, we will have $R>1$.

Using the method described in Sec.~\ref{subsec:floquet}, we now compute the limit-cycle solution for a wide range of couplings $\lambda$ and driving frequencies $\omega$, and look at both the rectification coefficient and the entropy production.  In Fig.~\ref{fig:rectification_coef} (a) we show the rectification coefficient as a function of  $\omega/\lambda$, for increasing values of $\lambda$. As can be seen, in the low and high frequency regimes the rectification coefficient goes to zero, but for intermediate values it peaks. 
For $\lambda > 1$, this peak occurs in the same value of $\omega/\lambda$.

In the limit cycle, the cycle-averaged entropy production rate is given simply by $\overline{\sigma} = \sum_\alpha \beta_\alpha \overline{J}^Q_\alpha$, where $\overline{J}^Q_\alpha$ denotes the cycle-averaged heat current associated with each bath. 
Interestingly, we have found that, even though there is rectification of energy current flowing into the left bath, the cycle-averaged entropy production rate, i.e., $\overline{\sigma}$ is the same in both the forward and reversed configurations. 
In Fig.~\ref{fig:rectification_coef} (b) we plot $\overline{\sigma}$ as a function of $\omega/\lambda$. 
We find that there is a strong peak in the entropy production rate at the same frequency in which the maximum rectification coefficient occurs. This clearly shows that an increase in rectification of energy current in our set-up comes at the cost of an increase in the entropy production rate.

We will like to mention that plots in Fig.~\ref{fig:rectification_coef}, which give the entropy cost of energy rectification as a function of driving frequency in our set-up, would be difficult to obtain via most other techniques. This is far beyond Markovian regime of system dynamics and additionally does not have any small parameter in the Hamiltonian to allow perturbation techniques. The frequency of the drive is also a free parameter, which has been varied from low to a high value, ruling out any possibility of perturbation in frequency also. This clearly highlights the power of our approach.

Our results in this section open many interesting questions, even for this rather simple set-up. The various interesting features, like the dependence of the position of the peak on various system parameters, the temperature dependence of rectification, the effect of having additional chemical potential bias, all can now be explored completely non-perturbatively. These directions deserve thorough separate investigations, which we delegate to future works.

\section{Conclusion}
\label{sec:conclusion}

In this work we have performed a detailed analysis of the mescoscopic leads in the context of quantum thermodynamics. Our main contribution is that the methodology allows for the inclusion of an arbitrary time dependence on the system Hamiltonian. This is a formidable challenge in quantum thermodynamics which we have demonstrated this technique can overcome. Following an overview of the thermodynamics of driven open quantum systems we reviewed the mesoscopic leads formalism and show how the definitions of the energy and particle current behave in the presence of driving. We emphasis that this technique has the formidable feature of being able to cope with both strong coupling to the leads and fast driving. We have demonstrated this focusing in particular on quadratic systems however an extension to interacting central systems by means of tensor networks is possible. We believe that the power of the methodology will allow for extensive explorations of thermodynamics of quantum systems in regimes that have so far been inaccessible with other approaches. To demonstrate the power of the methodology we apply our method to both the driven resonant level model and driven tunnel coupled quantum dot models respectively. In the case of the driven resonant level we were able to show that our results replicate known results in the literature in the high temperature limit. In the case of the driven quantum dot we showed how driving of one of the dot's energy can be used to induce a powerful heat rectification effect in a parameter regime which would be inaccessible to conventional techniques. In future work we plan to extend this technique further to access higher moments of the currents both in the presence and absence of driving as well as incorporating non quadratic interactions in the central system.

\textbf{Acknowledgements.}  The authors acknowledge Mark T. Mitchison for fruitful discussions. This work was funded by the European Research Council Starting Grant ODYSSEY (Grant Agreement No. 758403) and the EPSRC-SFI joint project QuamNESS. J.~G. is supported by a SFI-Royal Society University Research Fellowship. 
and GTL acknowledges the financial support of the S\~ao Paulo Funding Agency FAPESP (Grant No.~2019/14072-0.), and the Brazilian funding agency CNPq (Grant No. INCT-IQ 246569/2014-0). A.P acknowledges funding from the European Union’s Horizon 2020 research and innovation programme under the Marie Sklodowska-Curie Grant Agreement No. 890884. A.P also acknowledges funding from the Danish National Research Foundation through the Center of Excellence ``CCQ'' (Grant agreement no.: DNRF156). J.G. would aslo like to thank Rosario Fazio for suggestions regarding rectification in the presence of driving and M. T. Mitchison for general discussions on the project.

\section*{Appendix}

\appendix

\section{Brute force numerics with chain-mapping}
\label{appendix:chain mapping}
The systematic way to implement the brute force simulation of unitary dynamics of the system and baths using finite, but large enough, baths is offered by the chain mapping of the baths. Any bath spectral function, with finite upper and lower cut-offs, can be mapped into a semi-infinite one-dimensional nearest neighbour tight-binding chain with only the first site of the chain coupled with the system \cite{rc_mapping_QTD_book,Chain_mapping_bosons_fermions,rc_mapping_bosons,rc_mapping_fermions,Prior,Woods2015, Woods2016, Marscherpa2017, rc_mapping_old},
\begin{align}
& \hat{H}_{\alpha} = \sum_{p=1}^{L_B} \varepsilon_{p, \alpha} \hat{a}_{p, \alpha}^\dagger \hat{a}_{p, \alpha} + g_{p, \alpha} (\hat{a}_{p, \alpha}^\dagger \hat{a}_{p+1, \alpha} + \hat{a}_{p+1, \alpha}^\dagger \hat{a}_{p \alpha}), 
\nonumber \\
& \hat{H}_{S\alpha} = g_{0, \alpha} (S_\alpha^\dagger \hat{a}_{1, \alpha} + \hat{a}_{1,\alpha}^\dagger S_\alpha),  
\end{align}
with $L_B \to \infty$. Such a chain gives the spectral function $\mathcal{J}_{\alpha}(\omega)$ if  the on-site potentials $\varepsilon_{p,\alpha}$ and the hoppings $g_{p,\alpha}$ are obtained from the following set of recursion relations
\begin{align}
\label{rc_map}
& g_{p, \alpha}^2 = \frac{1}{2\pi} \int d\omega \mathfrak{J}_{p,\alpha}(\omega),\nonumber \\
&\varepsilon_{p+1, \alpha} = \frac{1}{2\pi g_{p, \alpha}^2} \int d\omega~\omega \mathfrak{J}_{p,\alpha}(\omega), \nonumber \\
&\mathfrak{J}_{p+1, \alpha}(\omega)= \frac{4g_{p,\alpha}^2 \mathfrak{J}_{p,\alpha}(\omega) }{\left[\mathfrak{J}_{p,\alpha}^H (\omega)\right]^2 +\left[\mathfrak{J}_{p,\alpha}(\omega)\right]^2},~~
\end{align}
with $p$ going from $0$ to $L_B$, $\mathfrak{J}_{0,\alpha}(\omega) =\mathcal{J}_{\alpha}(\omega) $ and $\mathfrak{J}_{p,\alpha}^H (\omega)$ being the Hilbert transform of  $\mathfrak{J}_{p,\alpha} (\omega)$,
\begin{align}
\mathfrak{J}_{p,\alpha}^H(\omega)= \frac{1}{\pi}\mathcal{P}\int_{-\infty}^{\infty} d\omega^\prime \frac{\mathfrak{J}_{p,\alpha}(\omega^\prime)}{\omega -\omega^\prime},
\end{align} 
where $\mathcal{P}$ denotes the principal value. 
With finite high and low frequency cut-offs, the parameters $\varepsilon_{p,\alpha}$ and $g_{p,\alpha}$ quickly tend to constants for increasing $p$. Let these constants be $\varepsilon_{B_\alpha}$ and $g_{B_\alpha}$. With baths modelled as such chains, the process to be simulated involves switching on the system-bath couplings at initial time. Due to Lieb-Robinson bounds, the information about this  spreads at a finite speed proportional to $g_{B_\alpha}$. 
Consequently, to simulate up to a time $t$, baths of size
\begin{align}
\label{eq:chain_mapping_bath_size}
L_B \sim g_{B,\alpha} t
\end{align}
suffices to accurately mimic the limit $L_B \to \infty$. We see that to simulate accurately up to a longer time, a larger bath size is required.

\bibliography{library}
\end{document}